\documentclass{llncs}
\usepackage[utf8]{inputenc}

\usepackage{Definitions}
\usepackage{eurosym}

\acrodef{SPHERE}{Sensor Platform for HEalthcare in Residential Environment}
\acrodef{EPSRC}{Engineering and Physical Sciences Research Council}
\acrodef{IRC}{Interdisciplinary Research Collaboration}
\acrodef{CASAS}{Centre for Advanced Studies in Adaptive Systems}
\acrodef{ADL}{Activities of Daily Living}
\acrodef{RGB}{Red-Green-Blue}
\acrodef{RGB-D}{RGB-Depth}
\acrodef{BLE}{Bluetooth Low Energy}

\acrodef{NTP}{network time protocol}
\acrodef{RSSI}{received signal strength indications}
\acrodef{PIR}{Passive Infra-Red}
\acrodef{ECML-PKDD}{European Conference on Machine Learning and Principles and Practice of Knowledge Discovery}
\acrodef{CSV}{Comma Separated Values}

\begin{document}

% 110 chars total
\title{The SPHERE Challenge}
\subtitle{Activity Recognition with Multimodal Sensor Data}
\titlerunning{The SPHERE Challenge}

\author{Niall Twomey\inst{1} \and Tom Diethe\inst{1} \and Meelis Kull\inst{1} \and Hao Song\inst{1} \and Massimo Camplani\inst{1} \and Sion Hannuna\inst{1} \and Xenofon Fafoutis\inst{1} \and Ni Zhu\inst{1} \and Pete Woznowski\inst{1} \and Peter Flach\inst{2} \and Ian Craddock\inst{1}}

\institute{Department of Electrical and Electronic Engineering, University of Bristol, UK,\\
\and
Department of Computer Science, University of Bristol, UK,\\
\Mundus~\url{http://irc-sphere.ac.uk/sphere-challenge/home}\\
\Letter~\email{firstname.lastname@bristol.ac.uk}}

\maketitle

\begin{abstract}
% Heterogeneous sensor fusion, multi-task, annotation uncertainty
This paper outlines the \ac{SPHERE} project and details the \ac{SPHERE} challenge that will take place in conjunction with \ac{ECML-PKDD} between March and July 2016. The \ac{SPHERE} challenge is an activity recognition competition where predictions are made from video, accelerometer and environmental sensors. Monitory prizes will be awarded to the top three entrants, with \euro 1,000 being awarded to the winner, \euro 600 being awarded to the first runner up, and \euro 400 being awarded to the second runner up. The dataset can be downloaded from the University of Bristol's data servers
\footnote{\url{https://data.bris.ac.uk/data/dataset/8gccwpx47rav19vk8x4xapcog}} \cite{spheredatadoi}.
\end{abstract}

\section{Background and Summary}
In this section we first describe the \ac{SPHERE} project, then the \ac{SPHERE} challenge is described, and outline of the remainder of the paper.

\subsection{The \ac{SPHERE} Project}
%(700 words maximum) An overview of the study design, the assay(s) performed, and the created data, including any background information needed to put this study in the context of previous work and the literature. The section should also briefly outline the broader goals that motivated the creation of this dataset and the potential reuse value. We also encourage authors to include a figure that provides a schematic overview of the study and assay(s) design. This section and the other main body sections of the manuscript should include citations to the literature as needed \cite{cite1, cite2}. References should be included within the manuscript file itself as our system cannot accept BibTeX bibliography files. Authors who wish to use BibTeX to prepare their references should therefore copy the reference list from the .bbl file that BibTeX generates and paste it into the main manuscript .tex file (and delete the associated \textbackslash{}bibliography and \textbackslash{}bibliographystyle commands).

Obesity, depression, stroke, falls, cardiovascular and musculoskeletal disease are some of the biggest health issues and fastest-rising categories of health-care costs. The financial expenditure associated with these is widely regarded as unsustainable and the impact on quality of life is felt by millions of people in the UK each day. 
Smart technologies can unobtrusively quantify activities of daily living, and these can provide long-term behavioural patterns that are objective, insightful measures for clinical professionals and caregivers. 

To this end the \acs{EPSRC}-funded ``\acf{SPHERE}'' \acf{IRC}\footnote{\url{http://www.irc-sphere.ac.uk/}} \cite{Woznowski15,Zhu15,diethe2015smart} has designed a multi-modal sensor system driven by data analytics requirements. The system is under test in a single house, and will be deployed in a general population of 100 homes in Bristol (UK).  The data sets collected will be made available to researchers in a variety of communities. 

Data is collected from the following three sensing modalities: 

\begin{enumerate}
    \item wrist-worn accelerometer; 
    \item RGB-D cameras (\ie video with depth information); and 
    \item passive environmental sensors. 
\end{enumerate}

\noindent With these sensor data, we can learn patterns of behaviour, and can track the deterioration/progress of persons that suffer or recover from various medical conditions. To achieve this, we focus activity recognition over multiple tiers, with the two main prediction tasks of \ac{SPHERE} including: 

\begin{enumerate}
    \item prediction of \ac{ADL} (\eg tasks such as meal preparation, watching television); and
    \item prediction of posture/ambulation (\eg walking, sitting, transitioning).
\end{enumerate}

Reliable predictions of \ac{ADL} allows us to model behaviour and of residents over time, \eg what does a typical day consist of, what times are particular activities performed \etc Prediction of posture and ambulation will complement \ac{ADL} predictions, and can inform us about the physical well-being of the participant, how mobile/responsive is the participant, how activie/sedintary, \etc

\subsection{The \ac{SPHERE} Challenge}
The task for the \ac{SPHERE} challenge is to predict posture and ambulation labels given the sensor data from the recruited participants. 

We will henceforth refer to posture/ambulation as `activity recognition' for brevity. It is worth noting that the term activity recognition has different interpretations when viewed from accelerometer, video, and environmental sensor perspectives. The definition of activities used in challenge most closely aligns to the terminology used in the field of accelerometer-based prediction. 

For this task, accelerometer, RGB-D and environmental data is provided. Accelerometer is samplled at 20 Hz and given in its raw format (see Section \ref{section:sensors:accel}). Raw video is not given in order to preserve anonymity of the participants. Instead, extracted features that relate to the centre of mass and bounding box of the identified persons are provided (see Section \ref{section:sensors:video}). Environmental data consists of \ac{PIR} sensors, and these is given in raw format (see Section \ref{section:sensors:env}). 

Twenty (posture/ambulation) activities labels are annotated in our dataset, and these are enumerated below together with short descriptions: 

\begin{multicols}{2}
    \begin{enumerate}
        \item \texttt{a\_ascend}: ascent stairs;
        \item \texttt{a\_descend}: descent stairs;
        \item \texttt{a\_jump}: jump; 
        \item \texttt{a\_loadwalk}: walk with load; 
        \item \texttt{a\_walk}: walk;
        \item \texttt{p\_bent}: bending; 
        \item \texttt{p\_kneel}: kneeling; 
        \item \texttt{p\_lie}: lying; 
        \item \texttt{p\_sit}: sitting; 
        \item \texttt{p\_squat}: squatting; 
        \item \texttt{p\_stand}: standing; 
        \item \texttt{t\_bend}: stand-to-bend; 
        \item \texttt{t\_kneel\_stand}: kneel-to-stand; 
        \item \texttt{t\_lie\_sit}: lie-to-sit; 
        \item \texttt{t\_sit\_lie}: sit-to-lie; 
        \item \texttt{t\_sit\_stand}: sit-to-stand;
        \item \texttt{t\_stand\_kneel}: stand-to-kneel; 
        \item \texttt{t\_stand\_sit}: stand-to-sit; 
        \item \texttt{t\_straighten}: bend-to-stand;  and
        \item \texttt{t\_turn}: turn
    \end{enumerate}
\end{multicols}

The prefix `\texttt{a\_}' on a label indicates an ambulation activity (\ie an activity requiring of continuing movement), the prefix `\texttt{p\_}' indicate static postures (i.e. times when the participants are stationary), and the prefix `\texttt{t\_}' indicate posture-to-posture transitions. These labels are the target variables that are to be predicted in the challenge. 

The \ac{SPHERE} challenge will take place in conjunction with the \ac{ECML-PKDD}\footnote{\url{http://www.ecmlpkdd2016.org}} conference, and will consist of two stages: 

\begin{itemize}
    \item \textbf{Stage 1. Scripted Data:} The first stage of the challenge uses sensor data that was recorded when the participants performed a pre-defined script. This script is described in later sections of this document. This data will be available from the challenge start to mid-April. 
    \item \textbf{Stage 2. Scripted + Real Data:} The scripted data from Stage 1 will be augmented with naturalistic data that was recorded from participants that did not follow a pre-defined script. Both train and test datasets will be augmented with this new data. This data will be available from mid-April to challenge end. 
\end{itemize}

Prizes will be awarded to the first three winners, with \euro 1,000 being awarded to the winner, \euro 600 to the runner up, and \euro 400 to the second runner up. Performance is evaluated using the weighted Brier score (see Section \ref{section:rules} for more details on performance evaluation). Winning participants must participate in the workshop that is associated with the challenge. Participation rules and eligibility criteria are detailed in Section \ref{section:rules}. 

% A number of features make this dataset interesting to activity recognition researchers: 
% \begin{enumerate}
%     \item We provide multimodal sensor data (i.e. sensor data from a plurality of sensing technologies) for activity recognition; and
%     \item Our data records are annotated by multiple annotators which will allow the participants to model annotators and their disagreement. 
% \end{enumerate}

\subsection{Structure of this Paper}
The remainder of this paper is structured as follows: in Section \ref{section:sensors} the set of sensors used in this dataset are described in greater detail, Section \ref{section:annotation} provides details regarding the annotation process, Section \ref{section:task} describes the primary task of this challenge in greater detail, Section \ref{section:data} provides a description of the format of the data, and in Section \ref{section:rules} a link to the full set of rules governing this eligibility for participation and winning this challenge is provided. 
% , data annotation, the task, and the data formats. The dataset can be obtained from \texttt{data.bris}\footnote{\url{http://data.bris.ac.uk/data}}.

%The vision of the SPHERE project  is to impact all these health-care needs simultaneously through data-fusion and pattern-recognition from a common platform of non-medical/environmental sensors at home.

\section{Sensors}\label{section:sensors}
The following subsections describe the sensing modalities that are found in the smart home. All sensors are synchronised with \ac{NTP}.

\subsection{Accelerometers}\label{section:sensors:accel}
Participants wore a device equipped with a tri-axial accelerometer on the dominant wrist, attached using a strap. The device wirelessly transmits data using the \ac{BLE} standard to several access points (receivers) positioned within the house. The outputs of these sensors are a continuous numerical stream of the accelerometer readings (units of $g$, \ie approximately \SI{9.81}{m.s^{-2}}). 
Accompanying the accelerometer readings are the \ac{RSSI} that were recorded by each access point (in units of dBm), and these data will be informative for indoor localisation. 
The accelerometers record data at \SI{20}{Hz}, and the accelerometer ranges are between \SI{\pm 8}{g}. \ac{RSSI} values are also recorded at \SI{20}{Hz}, and values are no lower than \SI{110}{dBm}. 

Due to the nature of the sensing platform, there may be missing packets from the data. Recent accelerometer work done by \ac{SPHERE} researchers on activity recognition with accelerometers includes \cite{diethe2016active,twomey2015bayesian}.

\subsection{\acs{RGB-D} Cameras}\label{section:sensors:video}
Video recordings were taken using ASUS Xtion PRO \ac{RGB-D} cameras\footnote{\url{https://www.asus.com/3D-Sensor/Xtion_PRO/}}. Automatic detection of humans was performed using the OpenNI library\footnote{\url{https://github.com/OpenNI/OpenNI}}, and false positive detections were manually removed by the organisers by visual inspection. Three \ac{RGB-D} cameras are installed in the \ac{SPHERE} house, and these are located in the living room, hallway, and the kitchen. No cameras are located elsewhere in the residence. 

In order to preserve the anonymity of the participants the raw video data are not shared. Instead, the coordinates of the 2D bounding box, 2D centre of mass, 3D bounding box and 3D centre of mass are provided. 

The units of 2D coordinates are in pixels (i.e. number of pixels down and right from the upper left hand corner) from an image of size $640\times480$ pixels. The coordinate system of the 3D data is axis aligned with the 2D bounding box, with a supplementary dimension that projects from the central position of the video frames. The first two dimensions specify the vertical and horizontal displacement of a point from the central vector (in millimetres), and the final dimension specifies the projection of the object along the central vector (again, in millimetres). 

\ac{RGB-D} data is very valuable in indoor environments as it can facilitate improvement in accuracy for fundamental computer vision tasks such as tracking \cite{massiTracking} while also enabling specific analysis at higher levels. \ac{RGB-D} data collected in the \ac{SPHERE} house has been used for example for specific action recognition \cite{actionPaper} and action quality estimation \cite{qualityPaper}.

\subsection{Environmental Sensors}\label{section:sensors:env}
The environmental sensing nodes are built on development platforms (Libelium, with CE marking)\footnote{\url{http://www.libelium.com/}}, powered by batteries or/and \SI{5}{V.DC} converted from mains. \ac{PIR} sensors are employed to detect presence in the data. Values of 1 indicate that motion was detected, whereas values of 0 mean that no motion was detected. A number of methods that deal with environmental sensor data are detailed in the following \cite{twomey2016unsupervised,twomey2014context}.

\subsection{Recruitment and Sensor Layout}

Participants were recruited to perform activities of daily living in the SPHERE house while the data recorded by the sensors was logged to a database.  Ethical approval was secured from the University of Bristol's ethics committee to conduct data collection, and informed consent was obtained from healthy volunteers. 

In the first stage of data collection, participants were requested to follow a pre-defined script\footnote{Details of the script can be found at \url{http://www.irc-sphere.ac.uk/sphere-challenge/script}}. In the second stage, participants were recorded and data was annotated in a non-scripted naturalistic setting. 

\begin{sidewaysfigure}
\centering
\begin{subfigure}{.5\textwidth}
  \centering
  \includegraphics[height=0.9\linewidth]{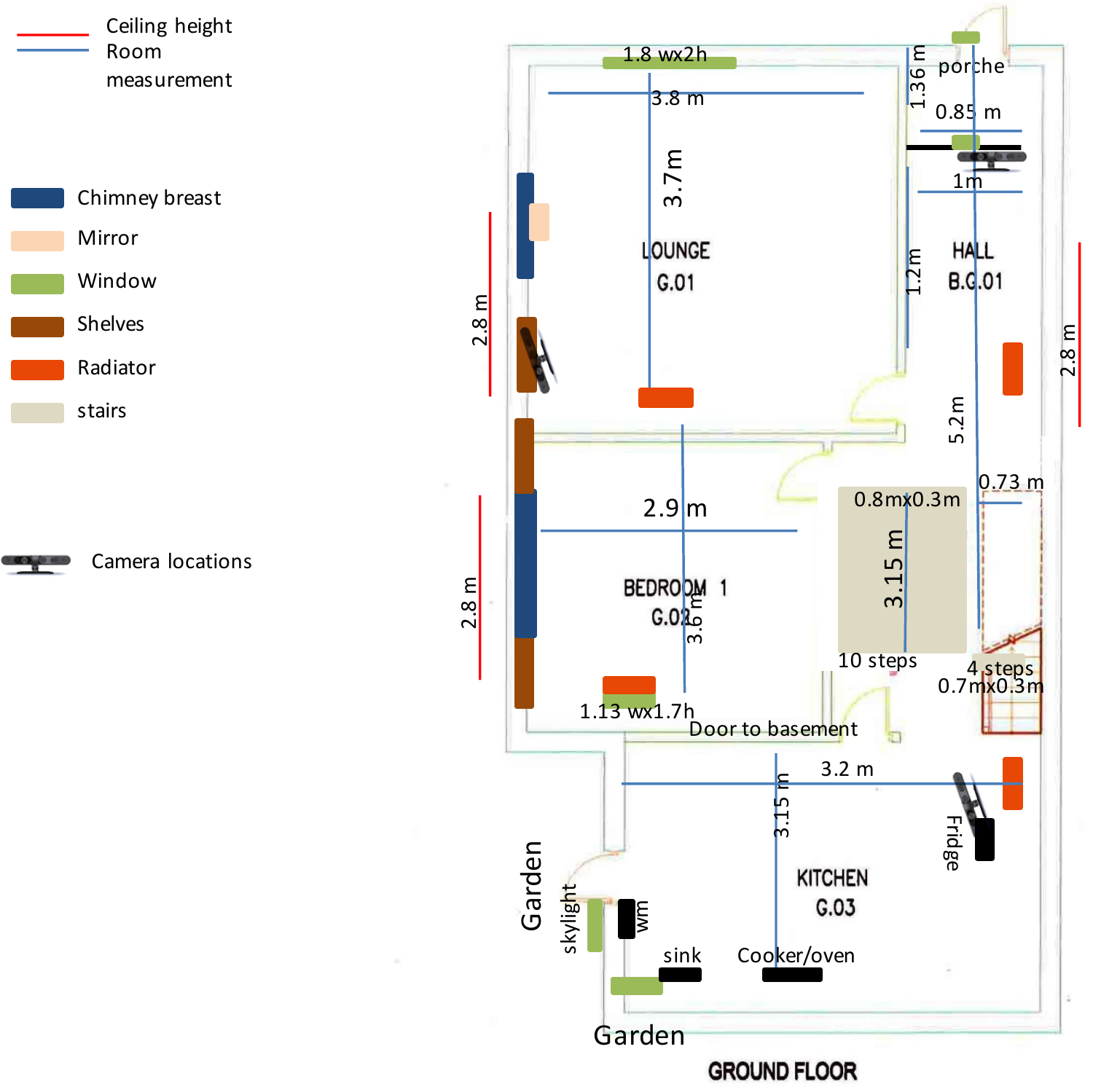}
  \caption{Ground floor}
  \label{fig:floorplan:0}
\end{subfigure}%
\begin{subfigure}{.5\textwidth}
  \centering
  \includegraphics[height=0.9\linewidth]{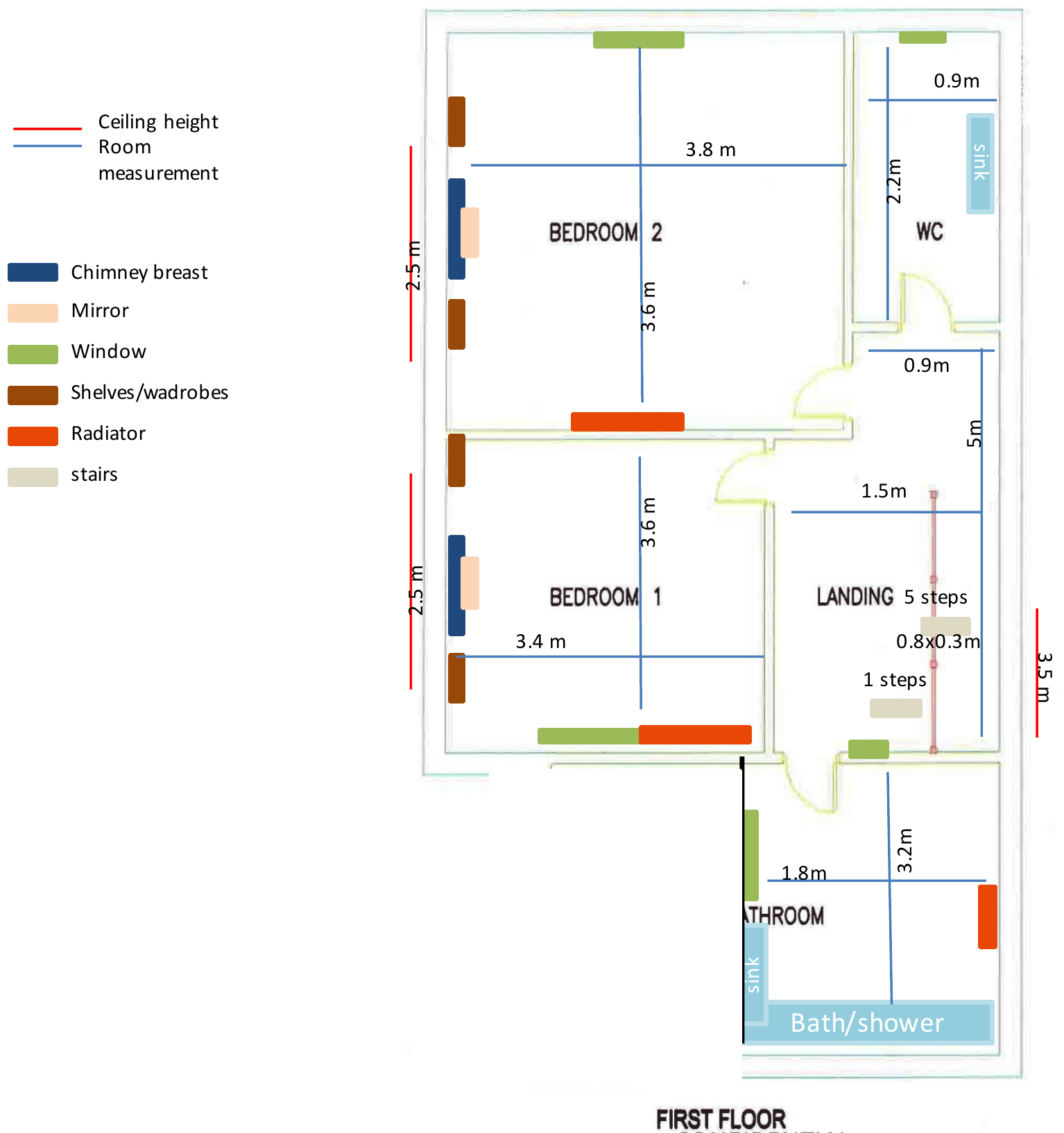}
  \caption{First floor}
  \label{fig:floorplan:1}
\end{subfigure}
\caption{The floorplan of the \ac{SPHERE} smart home.}
\label{fig:floorplan}
\end{sidewaysfigure}

Figures \ref{fig:floorplan:0} and \ref{fig:floorplan:1} show the floor plan of the ground and first floors of the smart environment respectively.

\section{Annotation}\label{section:annotation}
A team of 12 annotators were recruited and trained to annotate the set of activities outined in the introduction. 
%The classification task of this challenge is to predict the posture and ambulation of the participants as activities of daily living are performed. 
To support the annotation process, a head mounted camera (Panasonic HX-A500E-K 4K Wearable Action Camera Camcorder) recorded 4K video at 25 FPS to an SD-card. This data is not shared in this dataset, and is used only to assist the annotators. Synchronisation between the \ac{NTP} clock and the head-mounted camera was achieved by focusing the camera on an \ac{NTP}-synchronised digital clock at the beginning and end of the recording sequences. 

An annotation tool called ELAN\footnote{\url{https://tla.mpi.nl/tools/tla-tools/elan/}} was used for annotation. ELAN is a tool for the creation of complex annotations on video and audio resources, developed by the Max Planck Institute for Psycholinguistics in Nijmegen, The Netherlands.

Room occupancy labels are also given in the training sequences. While performance evaluation is not directly affected by room prediction on this, participants may find that modelling room occupancy may be informative for prediction of posture and ambulation.

\section{Task}\label{section:task}
The task is to predict posture, ambulation and transition activities for every second of the test data using the sensor data provided. Training data will be in the form of long sequences of data (tens of minutes in length), while test data will be short sequences, typically of length 10-30 seconds. Predictions must be made for every second of data. 

\subsection{Submission Files}

\texttt{sample\_submission.csv} contains an example submission file\footnote{This file is generated with: \url{https://github.com/IRC-SPHERE/sphere-challenge/blob/master/sample_submission.py}}. Each row should follow the \ac{CSV} format below. In this enumerated list, the the number indicates the column index of the \ac{CSV} file.  

\begin{multicols}{3}
    \begin{enumerate}
        \item test\_record\_id 
        \item start\_time
        \item end\_time 
        \item $Pr(a\_ascend)$
        \item $Pr(a\_descend)$
        \item $Pr(a\_jump)$
        \item $Pr(a\_loadwalk)$
        \item $Pr(a\_walk)$
        \item $Pr(p\_bent)$
        \item $Pr(p\_kneel)$
        \item $Pr(p\_lie)$
        \item $Pr(p\_sit)$
        \item $Pr(p\_squat)$
        \item $Pr(p\_stand)$
        \item $Pr(t\_bend)$
        \item $Pr(t\_kneel\_stand)$
        \item $Pr(t\_lie\_sit)$
        \item $Pr(t\_sit\_lie)$
        \item $Pr(t\_sit\_stand)$
        \item $Pr(t\_stand\_kneel)$
        \item $Pr(t\_stand\_sit)$
        \item $Pr(t\_straighten)$
        \item $Pr(t\_turn)$
    \end{enumerate}
\end{multicols}

Here, \texttt{test\_record\_id} is the name of the test directory, \texttt{start\_time} and \texttt{end\_time} are time stamps indicating the window over which prediction is to be made (note that the difference between \texttt{end\_time} and \texttt{start\_time} will always be 1 sec), and $Pr(.)$ is the predicted probability of each activity during the window. As an example, we would expect 10 predictions for a sequence of length 10 seconds.

\subsection{Performance Evaluation}
As both the targets and predictions are probabilistic, classification performance is evaluated with weighted Brier score\cite{brier1950verification}:

\begin{align}
    BS = \frac{1}{N} \sum_{n=1}^N \sum_{c=1}^C w_c \left( p_{n, c} - y_{n, c} \right)^2
\end{align}

\noindent 
where $N$ is the number of test sequences, $C$ is the number of classes, $w_c$ is the weight for each class, $p_{n, c}$ is the predicted probability of instance $n$ being from class $c$, and $y_{n, c}$ is the proportion of annotators that labelled instance $n$ as arising from class $c$. Lower Brier score values indicate better performance, with optimal performance achieved with a Brier score of $0$. 

For the first phase of the challenge (\ie with scripted data only) the class weights will be uniform, \ie $w_c=1 / C,~\forall~1 \leq c \leq C$. For the second phase (\ie with both scripted and real data) the weights will be adjusted to place more emphasis on the more difficult tasks. The changes to the weights will be announced on the challenge website and this document will be updated to reflect the changes. 

\section{Data}\label{section:data}
This section outlines the format of the data records available from this challenge. 

\subsection{Directory Structure}
Training data and testing data can be found in the `train' and `test' sub-directories respectively. The recorded data are collected under unique codes (each recording will be referred to as a `data sequence'). Timestamps are re-based to be relative to the start of the sequences, \ie each sequence always starts from $t=0$.

Each data sequence contains the following essential files: 
\begin{enumerate}
    \item \texttt{targets.csv}
    \item \texttt{pir.csv}
    \item \texttt{video\_hallway.csv}
    \item \texttt{video\_living\_room.csv}
    \item \texttt{video\_kitchen.csv}
    \item \texttt{meta.json}
\end{enumerate}

Two additional files are also provided in the training sequences: 
\begin{enumerate}
    \item \texttt{annotations\_*.csv}
    \item \texttt{location\_*.csv}
\end{enumerate}

The data from \texttt{annotations\_*.csv} is used to create the \texttt{targets.csv} file. \texttt{locations.csv} is available for participants that want to model the location. 

A \texttt{git} repository is publicly available \footnote{\url{https://github.com/IRC-SPHERE/sphere-challenge/}}. In this repository a number of scripts for visualisation, benchmarking and data processing are available. (All subsequent sensor images were generated using these scripts.)

\subsection{\texttt{targets.csv} (train only)}
% \label{sec:targets}
%
This file contains the probabilistic targets for classification. Multiple annotators may have annotated each sequence, and this file aggregates all of the annotations over 1 sec windows. These files can be created from \texttt{annotations\_*.csv}, and the following script shows how they are generated: \url{https://github.com/IRC-SPHERE/sphere-challenge/blob/master/gen_targets.py}. 

The following 20 activities are labelled:

\noindent
\begin{verbatim}
annotations = ('a_ascend', 'a_descend', 'a_jump', 'a_loadwalk', 
               'a_walk', 'p_bent', 'p_kneel', 'p_lie', 'p_sit', 
               'p_squat', 'p_stand', 't_bend', 't_kneel_stand', 
               't_lie_sit', 't_sit_lie', 't_sit_stand', 
               't_stand_kneel', 't_stand_sit', 't_straighten', 
               't_turn')
\end{verbatim}

\noindent where the prefix naming convention is the same as that described earlier. 

\texttt{targets.csv} files contain 22 columns: 

\begin{enumerate}
    \item \textbf{start}: The starting time of the window (relative to the start of the sequence)
    \item \textbf{end}: The ending time of the window (relative to the start of the sequence)
    \item \textbf{targets}: Columns 3-22: the 20 probabilistic targets. 
\end{enumerate}

\subsection{\texttt{pir.csv} (train + test)}

\begin{figure}[t]
  \centering
  \includegraphics[width=.95\linewidth]{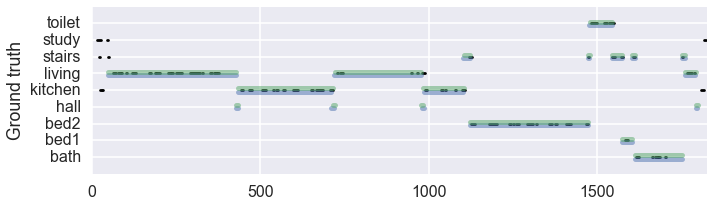}
    \caption{Example \ac{PIR} for training record \text{00001}. The black lines indicate the time durations where a \ac{PIR} is activated. The blue and green wide horizontal lines indicate the room occupancy labels as given by the two annotators that labelled this sequence.}
    \label{fig:example_pir}
\end{figure}

This file contains the start time and duration for all \ac{PIR} sensors in the smart environment. \ac{PIR} sensors are located in the following rooms:

\noindent
\begin{verbatim}
pir_locations = ('bath', 'bed1', 'bed2', 'hall', 'kitchen', 
                 'living', 'stairs', 'study', 'toilet')
\end{verbatim}

The columns of this CSV file are: 
\begin{enumerate}
    \item \textbf{start}: the start time of the \ac{PIR} sensor (relative to the start of the sequence)
    \item \textbf{end}: the end time of the \ac{PIR} sensor (relative to the start of the sequence)
    \item \textbf{name}: the name of the \ac{PIR} sensor being activated (from the \texttt{pir\_locations} list)
    \item \textbf{index}: the index of the activated sensor from the \texttt{pir\_locations}
\end{enumerate}

\noindent
Example \ac{PIR} sensor activations and ground truth locations are overlaid in \autoref{fig:example_pir}. Note, the \ac{PIR} sensors can be noisy in nature.

\subsection{\texttt{meta.json} (train + test)}
This file contains meta-data regarding the data records regarding the duration of the sequence. This data can be inferred from the sensor data but is provided for convenience.

\subsection{\texttt{acceleration.csv} (train + test)}

\begin{figure}[t!]
\centering
\begin{subfigure}{.90\textwidth}
  \centering
  \includegraphics[width=.95\linewidth]{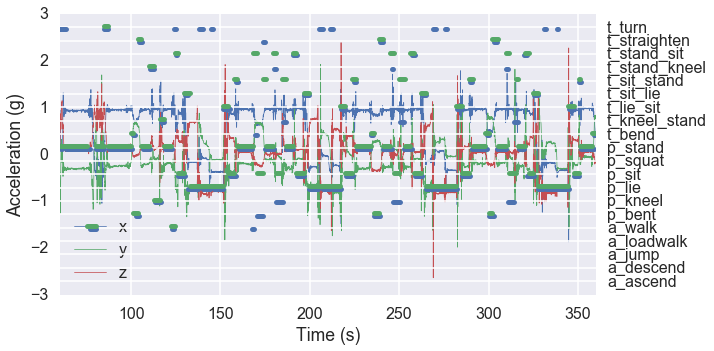}
  \caption{Acceleration signal trace shown over a 5 minute time period. Annotations are overlaid.}
  \label{fig:acceleration}
\end{subfigure}\\%
\begin{subfigure}{.90\textwidth}
  \centering
  \includegraphics[width=.95\linewidth]{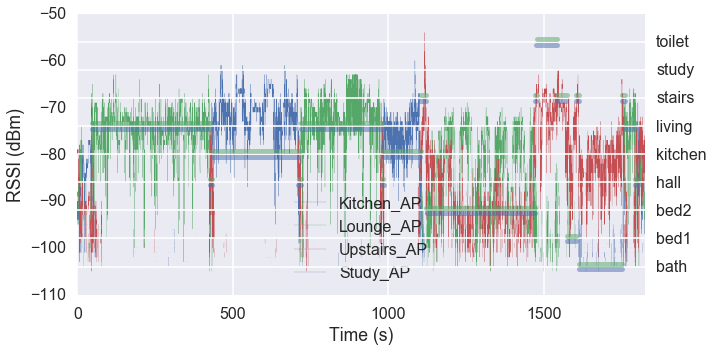}
  \caption{\ac{RSSI} values from the four access points. Room occupancy labels are shown by the the horizontal lines.}
  \label{fig:rssi}
\end{subfigure}
\caption{Example acceleration and \ac{RSSI} signals for training record \text{00001}. The line traces indicate the accelerometer/\ac{RSSI} values recorded by the access points. The horizontal lines indicate the ground-truth as provided by the annotators (two annotators annotated this record, and their annotations are depicted by the green and blue traces respectively).}
\label{fig:acceleration_rssi}
\end{figure}

The acceleration file consists of eight columns:

\begin{enumerate}
    \item \textbf{t}: this is the time of the recording (relative to the start of the sequence)
    \item \textbf{x}/\textbf{y}/\textbf{z}: these are the acceleration values recorded on the x/y/z axes of the accelerometer. 
    \item \textbf{Kitchen\_AP}/\textbf{Lounge\_AP}/\textbf{Upstairs\_AP}/\textbf{Study\_AP}: these specify the \ac{RSSI} of the acceleration signal as received by the access kitchen/lounge/upstairs access points. Empty values indicate that the access point did not receive the packet. 
\end{enumerate}

\noindent
Sample acceleration and \ac{RSSI} with overlaid annotations are shown in \autoref{fig:acceleration_rssi}. \autoref{fig:acceleration} shows the acceleration signals overlaid with activity labels, and \autoref{fig:rssi} shows the \ac{RSSI} signal information with the overlaid room occupancy.

\subsection{\texttt{video\_*.csv} (train + test)}

\begin{figure}[t]
  \centering
  \includegraphics[width=.95\linewidth]{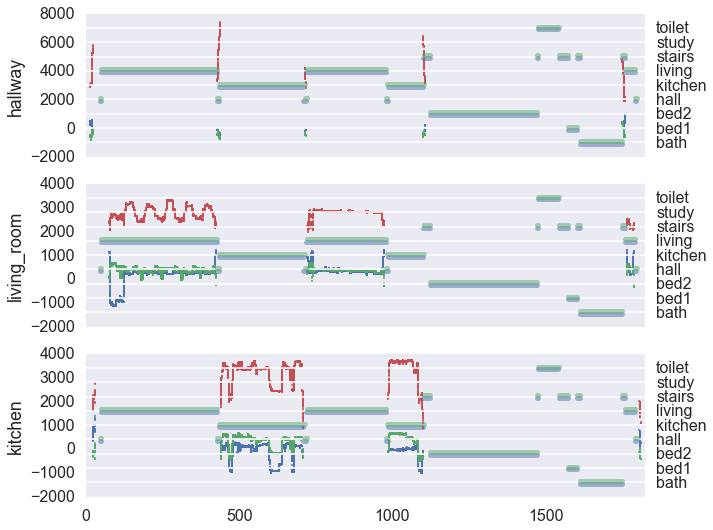}
    \caption{Example \texttt{centre\_3d} for training record \text{00001}. The horizontal lines indicate annotated room occupancy. The blue, green and red traces are the $x$, $y$, and $z$ values for the 3D centre of mass.}
    \label{fig:example_video}
\end{figure}

The following columns are found in the \texttt{video\_hallway.csv}, \texttt{video\_kitchen.csv} and \texttt{video\_living\_room.csv} files:
\begin{enumerate}
    \item \textbf{t}: The current time (relative to the start of the sequence)
    \item \textbf{centre\_2d\_x}/\textbf{centre\_2d\_y}: The x- and y-coordinates of the centre of the 2D bounding box. 
    \item \textbf{bb\_2d\_br\_x}/\textbf{bb\_2d\_br\_y}: The x and y coordinates of the bottom right (br) corner of the 2D bounding box 
    \item \textbf{bb\_2d\_tl\_x}/\textbf{bb\_2d\_tl\_y}: The x and y coordinates of the top left (tl) corner of the 2D bounding box
    \item \textbf{centre\_3d\_x}/\textbf{centre\_3d\_y}/\textbf{centre\_3d\_z}: the x, y and z coordinates for the centre of the 3D bounding box
    \item \textbf{bb\_3d\_brb\_x}/\textbf{bb\_3d\_brb\_y}/\textbf{bb\_3d\_brb\_z}: the x, y, and z coordinates for the bottom right back (brb) corner of the 3D bounding box
    \item \textbf{bb\_3d\_flt\_x}/\textbf{bb\_3d\_flt\_y}/\textbf{bb\_3d\_flt\_z}: the x, y, and z coordinates of the front left top (flt) corner of the 3D bounding box. 
\end{enumerate}

\noindent
Example 3D centre of mass data is plotted for the hallway, living room and kitchen cameras in \autoref{fig:example_video}. Room occupancy labels are overlaid on these, where we can see strong correspondence between the detected persons and room occupancy.

\subsection{Supplementary Files}

The following two sets of file need not be used for the challenge, but are included to facilitate users that wish to perform additional modelling of the sensor environment. For example, indoor localisation can be modelled with the \texttt{locations.csv} file, \cf \cite{fafoutis2015rssi}.

\subsubsection{\texttt{annotations\_*.csv}}  (train only):

This file provides the individual annotations as provided by the annotators. The target variables are the same as for \texttt{targets.csv}. % \ref{sec:targets}.
% The following 20 activities are annotated: 
%
% annotation\_names = ('a_ascend', 'a_descend', 'a_jump', 'a\_loadwalk', 'a_walk', 'p_bent', 'p_kneel', 'p_lie', 'p_sit', 'p_squat', 'p_stand', 't_bend', 't_kneel_stand', 't_lie_sit', 't_sit_lie', 't_sit_stand', 't_stand_kneel', 't_stand_sit', 't_straighten', 't_turn')
% As before, the prefix `a\_' indicates an ambulation activity (\ie an activity consisting of continuing movement), `p\_' annotations indicate static postures (\ie times when the participants are stationary), and `t\_' annotations indicate posture-to-posture transitions. 
%
The \texttt{annotations\_*.csv} contains the following four columns:

\begin{enumerate}
    \item \textbf{start}: the start time of the activity (relative to the start of the sequence)
    \item \textbf{end}: the end time of the activity (relative to the start of the sequence)
    \item \textbf{name}: the name of the label (from the \texttt{annotations} list)
    \item \textbf{index}: the index of the label name starting at 0
\end{enumerate}

\subsubsection{\texttt{location\_*.csv}}  (train only):

This file provides the annotation labels for room occupancy. The following rooms are labelled: 

\noindent 
\begin{verbatim}
location_names = ('bath', 'bed1', 'bed2', 'hall', 'kitchen', 
                  'living', 'stairs', 'study', 'toilet')
\end{verbatim}

\texttt{location.csv} contains the following four columns:

\begin{enumerate}
    \item \textbf{start}: the time a participant entered a room (relative to the start of the sequence)
    \item \textbf{end}: the time the participant left the room (relative to the start of the sequence)
    \item \textbf{name}: the name of the room (from the \texttt{location\_names} list)
    \item \textbf{index}: the index of the room name starting at 0
\end{enumerate}

\section{Eligibility and Rules}\label{section:rules}
The full set of rules for the challenge can be found at: 

\noindent
\url{http://irc-sphere.ac.uk/sphere-challenge/rules}
\section*{Acknowledgements}
This work was performed under the \ac{SPHERE} \ac{IRC} funded by the UK \ac{EPSRC}, Grant EP/K031910/1.

\end{document}